\documentclass[fleqn,12pt,twoside]{article}
\usepackage[headings]{espcrc1}

\readRCS
$Id: sorensen\_qm2005.tex,v 1.2 2004/02/24 11:22:11 spepping Exp $
\usepackage{graphicx}
\usepackage[figuresright]{rotating}

\newcommand{\AmS}{{\protect\the\textfont2
  A\kern-.1667em\lower.5ex\hbox{M}\kern-.125emS}}

\title{Identified Particle Spectra and Jet Interactions with the Medium}

\author{P. Sorensen\address[MCSD]{Brookhaven National Laboratory, 
    Upton, New York 11973-5000, USA} }%
       
\runtitle{Identified Particle Measurements}
\runauthor{P. Sorensen}

\begin{document}
\maketitle

\begin{abstract}
Through a combination of detector upgrades and advancements in
analysis techniques, physicist at RHIC are extending their ability to
identify particles at high $p_T$. With these new capabilities some
outstanding questions are being addressed: particularly regarding
baryon production. Here I present an overview of new data on
identified particle elliptic flow ($v_2$), identified particle $p_T$
spectra and identified hadron-hadron correlations. The observed
phenomena provide insights into hard parton interactions with the bulk
matter created in heavy-ion collisions and the relevant
degrees-of-freedom during its early stages.
\end{abstract}

\section{Introduction}

Physicists at the Relativistic Heavy Ion Collider (RHIC) have made
several unexpected observations~\cite{Adcox:2004mh}. Of particular
interest are measurements relating to baryon production in the
intermediate transverse momentum region ($1.5<p_T<5$
GeV/c)~\cite{Adler:2003kg,Adams:2003am,Sorensen:2004wg}. In
nucleon-nucleon collisions at $p_T=3$~GeV/c, one baryon is produced
for every three mesons (1:3). In $Au+Au$ collisions however, baryons and
mesons are created in nearly equal proportion (1:1). At this same
$p_T$, the elliptic anisotropy ($v_2$) of baryons is also 50\% larger
than meson elliptic flow: demonstrating that baryon production is also
enhanced in the direction of the impact vector between the colliding
nuclei
(in-plane)~\cite{Adams:2003am,Adams:2004bi}. This
enhancement in baryon production persists up to
$p_T=5.5$~GeV/c. Several possible explanations for the enhancement
are commonly considered:
\begin{quote} 
Multi-quark or gluon processes during hadron
formation---\emph{coalescence}~\cite{reco}.

Gluon configurations that carry baryon number---\emph{baryon
  junctions}~\cite{junctions,vitevjunctions}.

Collective motion amongst more massive baryons that populates the
higher $p_T$ regions of baryon $p_T$
spectra---\emph{flow}~\cite{hydro}.
\end{quote}
Coalescence models have garnered particular attention because they
seem to provide a natural explanation for the constituent-quark-number
scaling that has been observed in $v_2$ measurements. They also relate
hadronic observables to a pre-hadronic stage of interacting quarks and
gluons. As such, they touch on questions central to the heavy-ion
physics program: deconfinement and chiral symmetry restoration.

Given the potential physics benefits that can be derived from these
measurements, RHIC physicists have endeavored to extend their
abilities to identify hadron types up to higher $p_T$
regions~\cite{st_pid,ph_pid}. Through these efforts, measurements of
$\pi^{\pm}$, $\pi^{0}$, $K_S^0$, $\eta$, $p+\overline{p}$, and
$\Lambda+\overline{\Lambda}$ spectra and $v_2$ which extend beyond
$p_T = 8$~GeV/c have been presented at this conference. Here I review
those measurements along with correlation measurements that may help
clarify some of the questions remaining about the origin and
implications of the baryon enhancement in heavy-ion collisions.

\subsection{Azimuthal Dependence: $v_2$}

\begin{figure}[htb]
\centering\mbox{
\vspace{-10pt}
\includegraphics[width=0.8\textwidth]{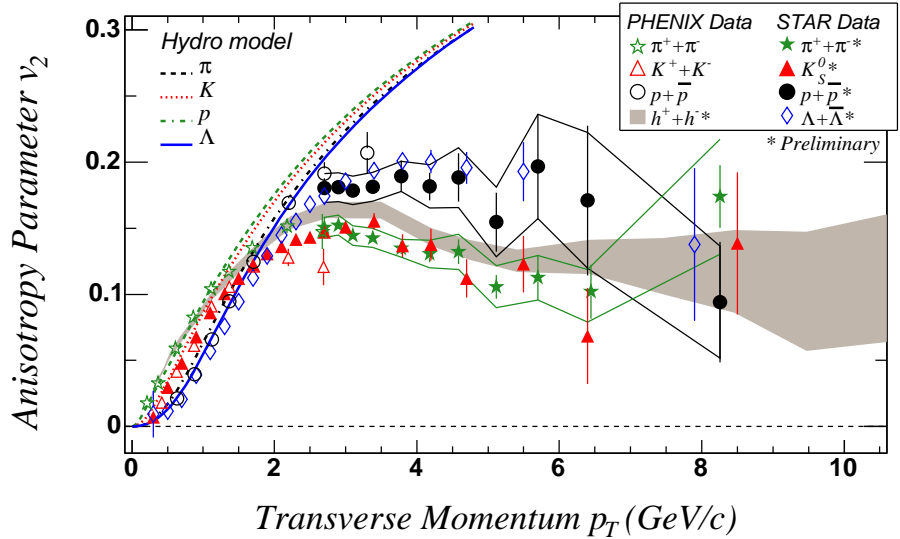}}
\vspace{-30pt}
\caption[]{ Elliptic flow measurements at middle rapidity from
  minimum-bias $Au+Au$ collisions at $\sqrt{s_{NN}}=200$~GeV. The bands
  around the STAR preliminary measurments of pions and protons
  represent systematic uncertainties mostly from non-flow
  correlations. The PHENIX measurements are made by correlating
  hadrons at middle rapdity with an event-plane measured using hadrons
  at $3.1<|\eta|<4.0$. For this reason they are less susceptible to most
  spurious, non-flow correlations. }
\label{v2}
\end{figure}

Fig.~\ref{v2} shows preliminary measurements of $v_2$ with a minimum
bias centrality selection in $Au+Au$ collisions at
$\sqrt{s_{_{NN}}}=200$~GeV~\cite{Oldenburg:2005er,Masui:2005aa}. The
curves show $v_2$ for pions, kaons, protons and Lambdas from a
hydrodynamic calculation~\cite{hydro}. At $p_T<1.0$~GeV/c, the mass
ordering demonstrates that $v_2$ in that region results from
collective motion. The hydrodynamic calculations capture the general
features of the data in this region. At much higher $p_T$ it is
expected that $v_2$ will be developed via energy loss by fast partons
as they traverse the medium created in the collisions. Calculations
suggest that $v_2$ from this mechanism should be less than
10\%~\cite{dedx}. It is also expected that all hadrons will have
similar $v_2$ values where energy loss mechanisms dominate $v_2$. The
magnitude and particle-type dependence of $v_2$ seen for $p_T<6$~GeV/c
suggest that other mechanisms contribute to the developement of $v_2$
up to 6--7 Gev/c. This observation is consistant with results from a
parton cascade model predicting that the effects of flow may persist
up to $p_T=7$~GeV/c~\cite{Molnar:flow}. At $p_T=7$~GeV/c, where,
within errors, the particle-type dependence of $v_2$ dissappears, the
$v_2$ measurements are consistant with expectations from energy loss
models~\cite{dedx}.

\begin{figure}[htbp]
\parbox{0.58\textwidth}{
\centering
\includegraphics[width=0.5\textwidth]{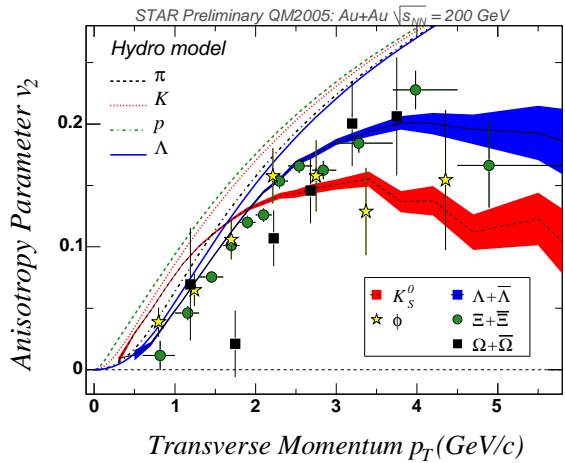}}
\parbox{0.42\textwidth}{
\caption[]{ Preliminary measurements of strange hadron elliptic flow
  shown by the STAR collaboration at QM2005. The measurements are made
  at middle-rapidity with a minimum-bias centrality selection
  corresponding to 0\%--80\% of the hadronic cross-section. Only
  statistical errors are shown. Curves are hydrodynamic model
  calculations~\cite{hydro}.}
\label{msbv2}
}
\end{figure}

Multi-strange hadron $v_2$ measurements with unprecedented accuracy
were presented at this conference~\cite{Oldenburg:2005er}. These
measurements are thought to be good tests for collective effects
amongst quarks and gluons~\cite{Adams:2005zg}. Measurements show that
when the $p_T$ spectra of $\phi$-mesons and $\Omega$-baryons are
characterized with a two component hydrodynamic inspired fit, the
temperature and flow velocity parameters extracted are systematically
different than those extracted for non-strange
hadrons~\cite{Adcox:2004mh}. The differences are consistent with less
hadronic rescattering amongst multi-strange hadrons. For this reason,
it is believed that the $v_2$ of multi-strange hadrons reflect
collectivity that developed before hadrons formed.

In Fig.~\ref{msbv2} $v_2$ measurements for $K_S^0$, $\phi$, $\Lambda +
\overline{\Lambda}$, $\Xi+\overline{\Xi}$, and
$\Omega+\overline{\Omega}$ from minimum bias $Au+Au$ collisions at
$\sqrt{s_{_{NN}}}=200$~GeV are shown. The $\phi$ and $\Omega$
$v_2$ values follow the same systematic trends as the non-strange
hadrons: with an apparent mass ordering at low $p_T$ and a
quark-number dependence at intermediate $p_T$. Combined with the
spectra measurements, these measurements suggests that $v_2$ is
developed during a prehadronic stage.

\subsection{Quark Number Dependence}

\begin{figure}[htb]
\parbox{0.55\textwidth}{
\centering
\includegraphics[width=0.55\textwidth]{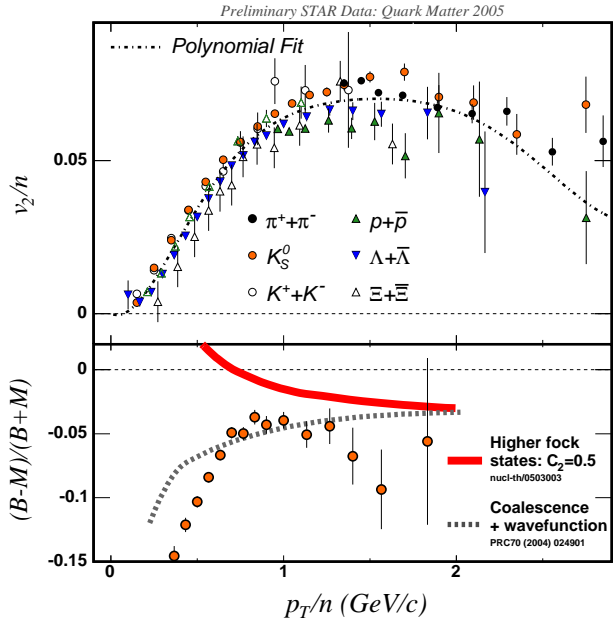}
\vspace{-50pt}
}
\parbox{0.45\textwidth}{
\caption[]{ Top panel: Quark number scaled elliptic flow for
  identified hadrons. A polynomial is fit to all data points. Bottom
  panel. The difference between quark number scaled baryon $v_2$ and
  quark number scaled meson $v_2$ divided by the sum:
  $(B-M)/(B+M)$. Here the ratio is taken using hyperons and
  kaons. The solid curve shows model predictions when the daughter
  partons of the constituent quarks are considered. The dashed line
  shows model results after the momentum distributions of the
  constituent quarks are taken into account. }
\label{nqscaling}}
\end{figure}

Hadronization by coalescence or recombination of constituent quarks is
thought to explain many features of hadron production in the intermediate
$p_T$ region ($1.5<p_T<5$~GeV/c)~\cite{reco}. These models find that
at intermediate $p_T$, $v_2$ may follow a quark-number ($n_q$) scaling
with $v_2(p_T/n_q)/n_q$ for all hadrons falling on one curve. This
scaling behavior has been observed in $Au+Au$ collisions at 200
GeV~\cite{Adams:2004bi}. More sophisticated theoretical considerations
have led to predictions of fine structure in quark-number scaling---with
predictions for a baryon $v_2/n_q$ being smaller than meson
$v_2/n_q$~\cite{Greco:2004ex,fock}.

Fig.~\ref{nqscaling} (top panel) shows $v_2$ vs $p_T$ for identified
particles, where $v_2$ and $p_T$ have been scaled by $n_q$. A
polynomial function is fit to the scaled values. The deviations from
the scaling are shown in the bottom panel by plotting the difference
between the scaled baryon $v_2$ and the scaled meson $v_2$ divided by
the sum $(B-M)/(B+M)$. Theoretical predictions for this value are also
shown. The model that compares best to data is a coalescence model
that includes the effect of quark momentum distributions inside the
hadron (\textit{Coalescence + wavefunction})~\cite{Greco:2004ex}. The
effect of accounting for the substructure of constituent quarks
(\textit{higher fock states})~\cite{fock} leads to a negative
$(B-M)/(B+M)$ ratio but with a smaller magnitude than observed in the
data. It should be noted, however, that the systematic
uncertainties on the measurements have not yet been estimated.

\begin{figure}[htb]
\vspace{-10pt}
\centering\mbox{
\includegraphics[width=1.0\textwidth]{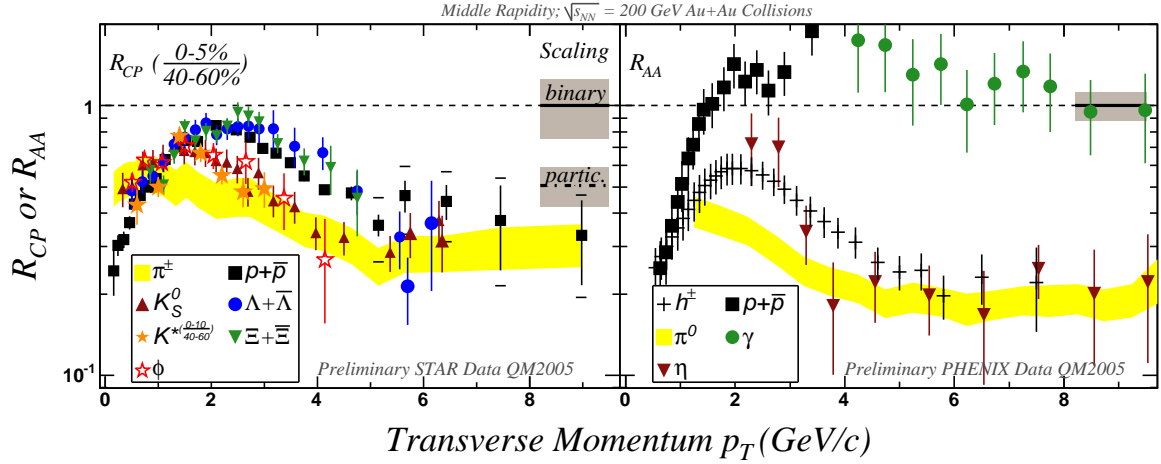}}
\vspace{-30pt}
\caption[]{ Preliminary identified particle $R_{CP}$ (left panel) and
  $R_{AA}$ (right panel). Grey bands represent the error on the
  $N_{bin}$ and $N_{part}$ calculations. For the $R_{AA}$ data, error
  bars represent both systematic and statistical uncertainties. For
  pion $R_{CP}$, the yellow band includes both systematic and
  statistical uncertainties. For proton $R_{CP}$, the systematic
  uncertainties are shown as brackets on the final five points
  (systematic uncertainties are similar for lower $p_T$ points). }
\label{rcp}
\end{figure}

Quark-number dependences at intermediate $p_T$ are also manifested in
the centrality dependence of the $p_T$ spectra. Fig.~\ref{rcp} shows
the ratios $R_{CP}$~\cite{Adams:2003am,RCP,olga,dunlop} (central
$Au+Au$ over peripheral $Au+Au$) in the left panel and
$R_{AA}$~\cite{phenix_spec,ph_pid_qm} ($Au+Au$ over $p+p$) in the
right panel where the spectra have been scaled by the number of binary
nucleon-nucleon collisions ($N_{bin}$). Hard processes are expected to
scale with $N_{bin}$, so $R_{AA,CP}<1$ at high $p_T$ indicates a
suppression due to nuclear affects. Models incorporating energy loss
of hard scattered partons in medium successfully describe the
suppression of hadrons at high $p_T$~\cite{dedx}. Indeed, photons
which only interact electro-magnetically are found to follow $N_{bin}$
scaling. Here, however, we concentrate on the features of the
particle-type dependence of hadrons at intermediate $p_T$.

For $p_T>1.5$~GeV/c, $R_{CP}$ values are grouped by particle type
(mesons vs. baryons). The larger baryon $R_{CP}$ values indicate that
baryon production increases more quickly with centrality than meson
production. This observation is confirmed with good precision for
protons and hyperons. The $\phi$ is a particularly interesting test
particle since it is a meson but is more massive than the proton. With
good precision, the $\phi$ is now confirmed to follow the systematics
of the other mesons. Additionally $R_{AA}$ has been measured for the
$\eta$-meson and its values are similar to those of the
pions.

Particular progress has been made in the measurement of charged pion
and proton $R_{CP}$. These measurements now extend into the region
where fragmentation in vacuum is thought to be the dominate
hadronization mechanism. In this region, if the relative fraction of
quark and gluon jets (fragmenting to a hadron with a given $p_T$)
changes from peripheral to central $Au+Au$ collisions, the $R_{CP}$ of
different particle types may take on different values. This is a
result of the difference in the shapes of the fragmentation functions
of quark and gluon jets. Within errors however, the $\pi^{\pm}$,
$K_S^0$, $p+\overline{p}$, and $\Lambda+\overline{\Lambda}$ $R_{CP}$
values at $p_T=6.5$~GeV/c appear to be consistent with each other.
The systematic errors on the proton $R_{CP}$ still, however, allow for
significant deviations between proton and pion $R_{CP}$.

Measurement of $K_S^0$ and $\Lambda+\overline{\Lambda}$ $R_{CP}$ at
$\sqrt{s_{_{NN}}}=17.2$~GeV indicate that baryon production is also
enhanced at lower center-of-mass energies~\cite{Dainese:2005vk}: with
$\Lambda+\overline{\Lambda}$ $R_{CP}$ values well above $N_{bin}$
scaling. It has been proposed that the large baryon $R_{CP}$ values at
RHIC may be a result of the suppression of meson production coupled
with novel baryon production mechanisms scaling with $N_{bin}$ or
$N_{part}$~\cite{vitevjunctions}. The lower energy measurements call
this interpretation into question since baryon production is still
enhanced even while the suppression of mesons is much weaker.

\begin{figure}[htbp]
\parbox{0.55\textwidth}{
\centering
\includegraphics[width=0.55\textwidth]{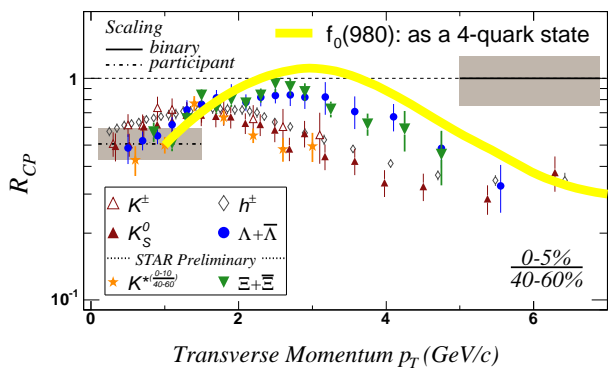}
\vspace{-40pt}
}
\parbox{0.45\textwidth}{\caption[]{ A sketch of what $R_{CP}$ may look
    like for the $f_{0}$(980) meson if it is a non-$q\overline{q}$
    state. At $p_T=3$~GeV, $f_0$(980) $R_{CP}>1$ would favor the
    $f_0$(980) as a four-quark or $K\overline{K}$-molecular state. If
    $f_0$(980) $R_{CP}$ values are similar to those for other mesons,
    either the $f_0$(980) is a two-quark state or
    coalescence/recombination is not the dominant formation
    mechanism.}
\label{f0}}
\end{figure}

Since the grouping of $R_{CP}$ with quark number appears to extend
across many hadron species, the centrality dependence of intermediate
$p_T$ particle yields may provide information about the quark content
of non-$q\overline{q}$ candidates. The $f_0$(600), $f_0$(980),
$a_0$(980), and $\kappa$(800) are all candidates for four-quark or
$K\overline{K}$-molecular states~\cite{pdg,f0theory,f0theory2}. The
branching ratio for $\phi \rightarrow \pi^0\pi^0\gamma$ decays, which
is sensitive to the structure of the $f_0$, favors the $f_0$ as a
four-quark state~\cite{f0exp}. In Fig.~\ref{f0}, a sketch is shown of
what $R_{CP}$ may look like for the $f_0$ if it is a four-quark or
$K\overline{K}$-molecular state.
$f_0$(980) $R_{CP}$ can be measured from the $\pi^+\pi^-$ invariant
mass spectrum~\cite{Adams:2003cc}.  If at $p_T=3$~GeV/c its value is
similar to other mesons, the $f_0$ is either a two quark-state or the
baryon enhancement is not a function of the number of quarks: strongly
favoring baryon junctions as the source of the baryon enhancement. The
result that $f_0$ $R_{CP}$ is greater than the baryon $R_{CP}$, will
strongly favor $f_0$ as a four-quark or $K\overline{K}$-molecular
state and coalescence as the dominant hadronization process for
intermediate $p_T$ hadrons.  Measurements of the $\pi^+\pi^-$
invariant mass spectrum should be sensitive to the large difference
between these scenarios.

\section{Baryon to Meson Ratios}

\begin{figure}[htb]
\centering\mbox{
\vspace{-15pt}
\includegraphics[width=1.0\textwidth]{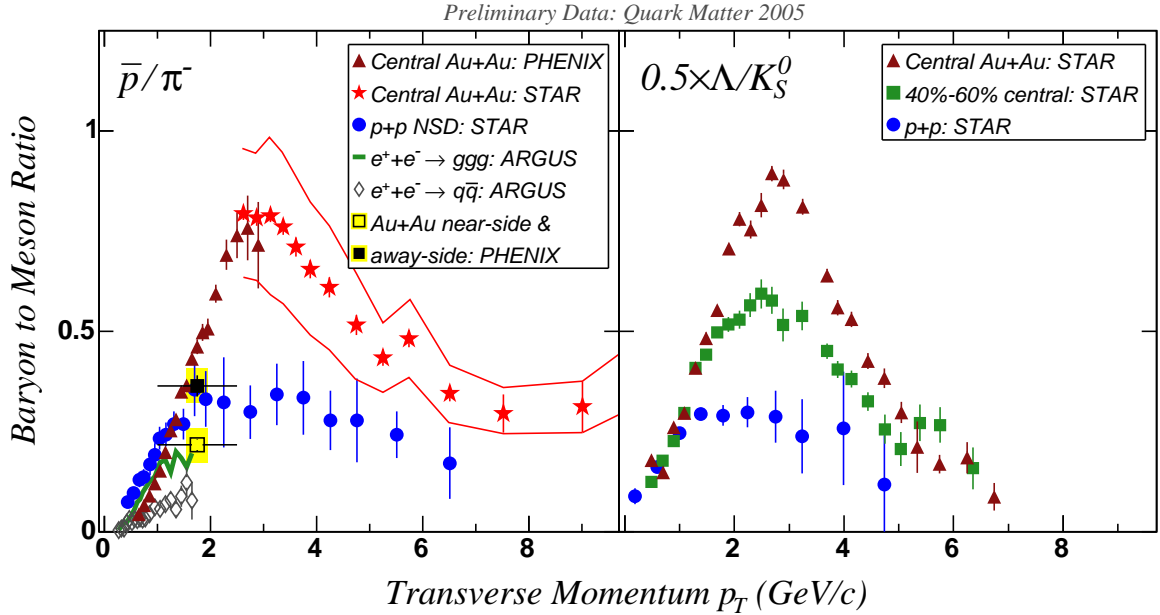}}
\vspace{-35pt}
\caption[]{ Left panel: the $\overline{p}/\pi^-$ ratio at middle
  rapidity for central $Au+Au$ and $p+p$ collisions at
  $\sqrt{s_{_{NN}}}=200$~GeV. ARGUS measurements of the proton to pion
  ratio in $e^++e^-$ collisions at $\sqrt{s}=10$~GeV are shown for two
  classes of events: $e^+e^- \rightarrow \Upsilon \rightarrow ggg$ and
  continuum events dominated by $e^+e^- \rightarrow
  q\overline{q}$. Measurements of the proton to pion ratio made for
  particles associated with a trigger hadron ($p_T>2.5$) are also
  shown. Left panel: $\Lambda/K_S^0$ in central $Au+Au$,
  mid-peripheral $Au+Au$ and minimum-bias $p+p$ collisions. Values are
  scaled by 0.5.}
\label{ratios}
\end{figure}

The quark-number scaling of $v_2$ favors the number of
constituent-quarks as the relevant quantity in the baryon
enhancement. Before this scaling was observed, however, the baryon
enhancement was seen in the proton-to-pion
ratio~\cite{Adler:2003kg}. Fig.~\ref{ratios} (left panel) shows the
$\overline{p}/\pi^-$ ratio measured in $e^++e^-$~\cite{argus},
$p+p$~\cite{olga}, and $Au+Au$~\cite{Adler:2003kg,olga}
collisions. The right panel shows the $\Lambda/K^0_S$ ratio for $p+p$,
mid-peripheral $Au+Au$, and central $Au+Au$ collisions scaled by 0.5
to fit on the same axis. The baryon enhancement at intermediate $p_T$
is pronounced in central $Au+Au$ collisions with $\overline{p}/\pi^-$
reaching a maximum value of nearly 1 at $p_T\approx3$~GeV/c. The
baryon junction calculations in Ref.~\cite{vitevjunctions} predict
that the $p_T$ value where the $B/M$ ratio is at its maximum will
increase with collision centrality. This prediction can be compared to
the $\Lambda/K_S^0$ data. Measurements are still not precise enough,
however, to confirm nor disprove this prediction.

Fig.~\ref{ratios} also demonstrates that the baryon enhancement in
$Au+Au$ collisions may be part of a systematic trend.  Baryon
production is also enhanced in $\sqrt{s}=10$~GeV $e^++e^-$ collisions
when $e^+e^- \rightarrow \Upsilon \rightarrow ggg$ events are compared
to continuum $e^+e^- \rightarrow q\overline{q}$
events~\cite{argus}. From those measurements the question arises:
\textit{is the enhancement related to multi-parton topological effects or a
difference between quark and gluon fragmentation?} Since $e^+e^-
\rightarrow \Upsilon \rightarrow ggg$ is a purely gluonic process, the
observation that the $\overline{p}/\pi^-$ ratio is even larger in
$Au+Au$ collisions indicates that multi-parton topological effects
drive the enhancement (baryon junctions or coalescence).

\section{Jet Interactions with the Medium}

Also included in the left panel of Fig.~\ref{ratios} are measurements
of the $(p+\overline{p})/(\pi^++\pi^-)$ ratio in jets observed in
$Au+Au$ collisions~\cite{Sickles:2004jz}. The analysis triggers on
``high $p_T$" hadrons ($2.5<p_T<4.0$~GeV/c) and measures the proton
and pion yields in the peak around the trigger hadron
(near-side) and the peak opposite the trigger hadron
(away-side). Because of energy-loss, triggering on high $p_T$ hadrons
will select events where a hard scattering occurred near the surface of
the $Au+Au$ collision overlap region. In this case, one of the fast
partons can escape relatively unperturbed while the other fast parton
must traverse the bulk matter created in the $Au+Au$ collision. The
near-side baryon-to-meson ratio then should reflect hadron production
outside the matter, while the away-side ratio should reflect hadron
production within the matter. The result
$(p+\overline{p})/(\pi^++\pi^-)_{away-side} >
(p+\overline{p})/(\pi^++\pi^-)_{near-side}$ is then consistent with
the baryon enhancement arising from multi-gluon or multi-quark
processes in hadron production.

\begin{figure}[htb]
\vspace{-5pt}
\hspace{0.1\textwidth}
\resizebox{0.33\textwidth}{!}{\includegraphics{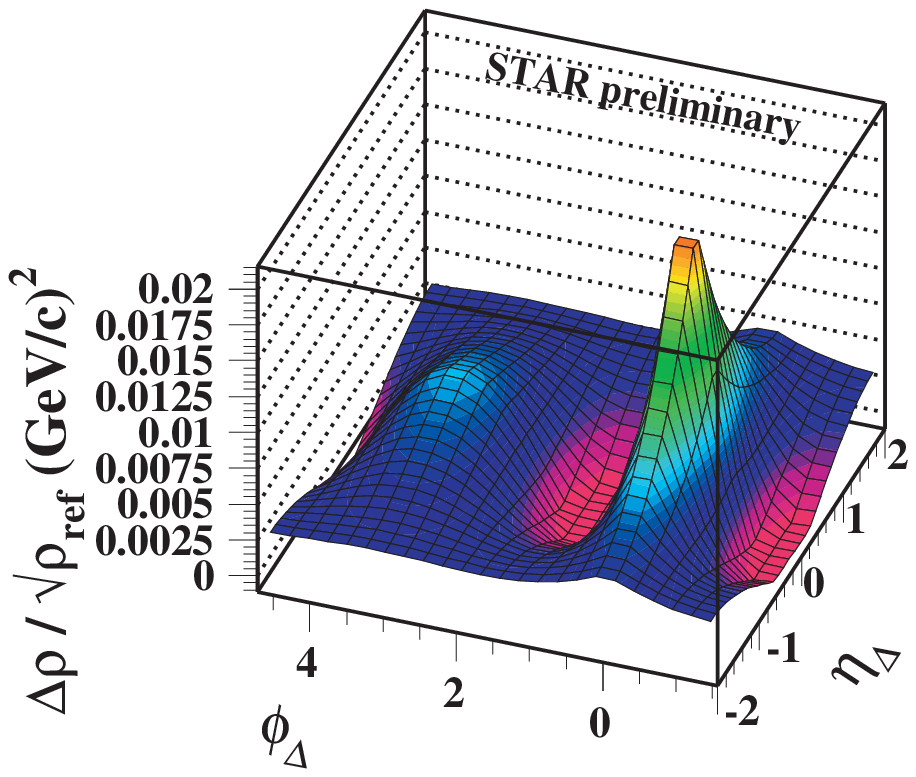}}
\hspace{0.15\textwidth}
\resizebox{0.22\textwidth}{!}{\includegraphics[width=0.5in,height=0.4in,angle=90]{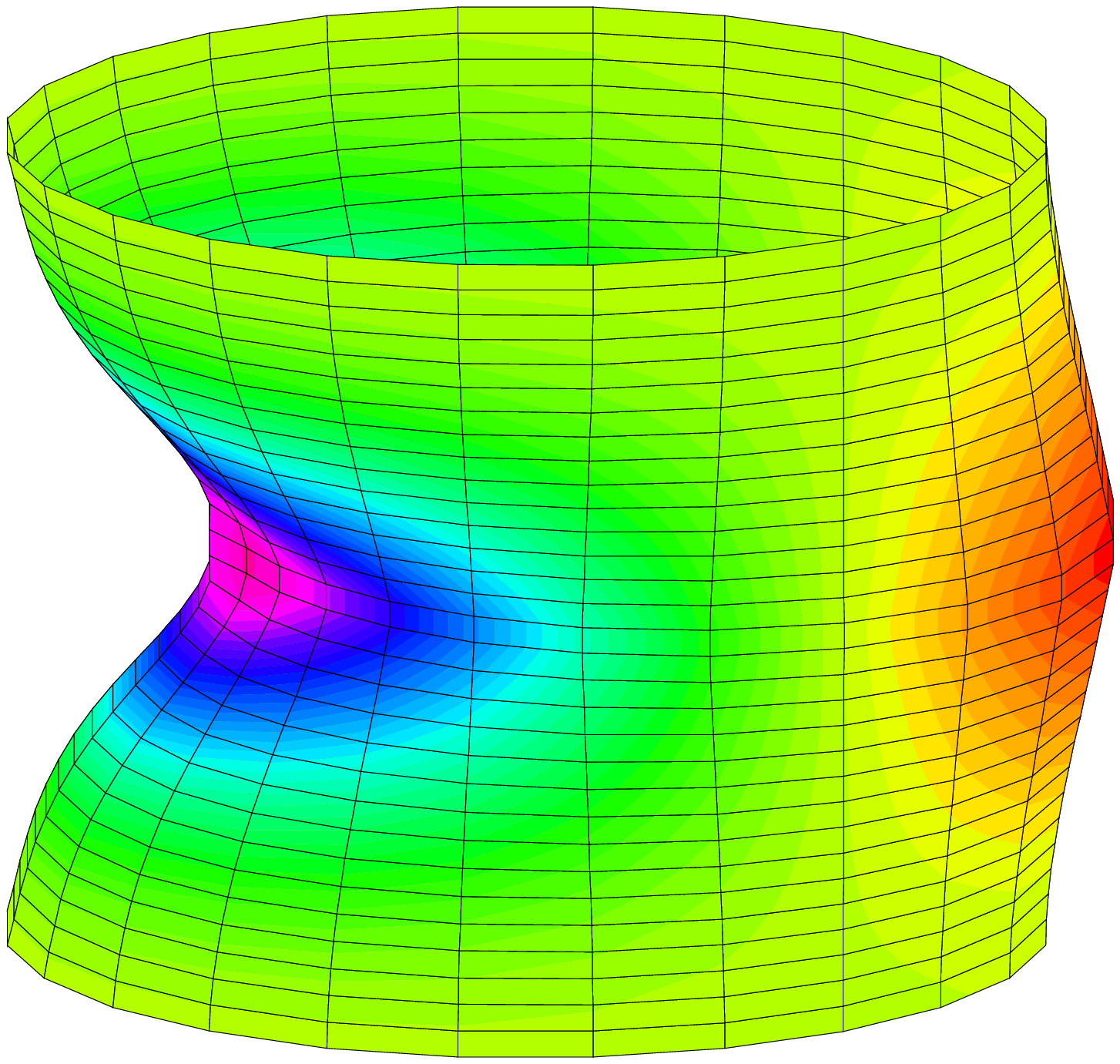}}
\vspace{-15pt}
\caption[]{ Left panel: $p_T$ autocorrelations derived from the
  $\langle p_T \rangle$ fluctuation scale dependence in $Au+Au$
  collisions at $\sqrt{s_{_{NN}}} = 200$
  GeV~\cite{Adams:2005aw}. Sinusoidal modulations associated with
  $v_2$ have been subtracted. Right panel: $p_T$ autocorrelations
  plotted in cylindrical coordinates. The positive \textit{near-side}
  mini-jet peak is subtracted. The figure demonstrates that the medium
  in $Au+Au$ collisions recoils in response to an impinging
  parton~\cite{Trainor:2005ac}.}
\label{autocorr}
\end{figure}

Baryon production may be further favored as a result the recoil that
the matter undergoes in response to an impinging hard
parton. Fig.~\ref{autocorr} shows measurements of $p_T$
autocorrelations in 20\%--30\% central, 200 GeV, $Au+Au$
collisions~\cite{Adams:2005aw,Trainor:2005ac}. The left panel shows
the autocorrelations after subtracting sinusoidal modulations
associated with $v_2$. The remaining features are a positive near-side
peak (associated with an escaping hard parton), a negative near-side
valley, and a broad, positive, away-side peak. The right panel shows
the same data plotted on a cylinder after the positive near-side peak
is subtracted. The data show that the $p_T$ distribution of
hadrons---close in $\eta$ and $\phi$ to an escaping hard parton---is
red-shifted. This apparent recoil can lead to collective motion
amongst the matters constituents in the direction of the impinging
partons momentum. This jet-induced flow will lead to an increase in
the number of co-moving constituents on the away-side of an escaping
hard parton. The data, in Figs.~\ref{v2}, \ref{rcp}, and~\ref{ratios}
show that wherever the density of co-moving constituents is larger,
baryon production is enhanced. Now, the near-side and away-side data
in~\ref{ratios} shows that baryon production is also enhanced in the
recoil region of quenched jets.

\section{Anti-baryon to Baryon Ratios}

\begin{figure}[htb]
\vspace{-10pt}
\centering\mbox{ \includegraphics[width=1.0\textwidth]{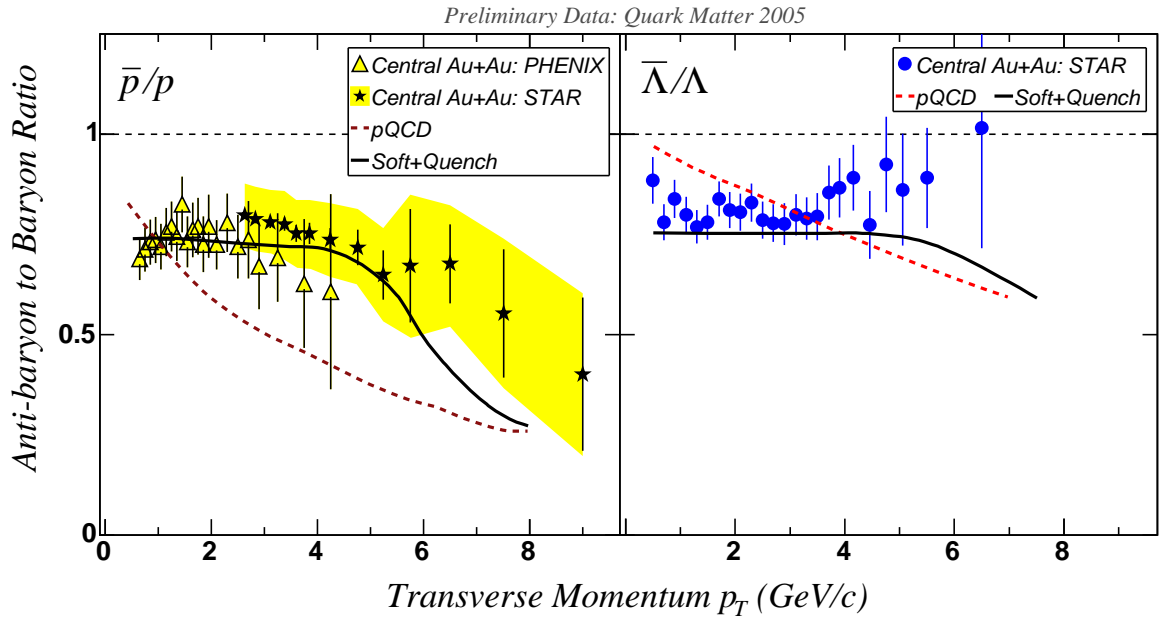}}
\vspace{-30pt}
\caption[]{ Left panel: $\overline{p}/p$ at middle rapidity from
  central $Au+Au$ collisions for $\sqrt{s_{_{NN}}}=200$~GeV. The yellow
  band represents the systematic uncertainties on the preliminary STAR
  measurements. Right panel: Preliminary STAR measurements of
  $\overline{\Lambda}/\Lambda$ at middle rapidity in central $Au+Au$
  collisions at $\sqrt{s_{_{NN}}}=200$~GeV.}
\label{bbarb}
\end{figure}

Further information about baryon production can be gleaned from
anti-baryon to baryon ratios ($\overline{B}/B$). Fig.~\ref{bbarb}
shows $\overline{B}/B$ for protons (left panel) and lambda hyperons
(right panel). The curves show model expectations from
Ref.~\cite{vitevjunctions}. Disagreement between the pQCD calculation
and the data are evident and indicate the importance of
non-perturbative effects in baryon production at $p_T<6$~GeV/c. The
\textit{Soft+Quench} calculation uses a phenomenological treatment of
baryon junctions to successfully describe the observed trends. For
this calculation, the apparent baryon enhancement is revealed through
the suppression of high-$p_T$ meson production. It is not clear,
however, that this model can successfully describe the
$\sqrt{s_{_{NN}}}$, and system-size dependence of the
$\overline{p}/\pi^-$ ratio: particularly given that baryon production
appears to be enhanced in $d+Au$ collisions~\cite{Ruan:2005hy}
collisions and $Pb+Pb$ collisions at
$\sqrt{s_{_{NN}}}=17.2$~GeV~\cite{Dainese:2005vk}, where little or no
high-$p_T$ suppression is seen.

\section{Conclusions}

Pion, kaon, proton, and hyperon momentum-space distributions have been
measured up to $p_T \approx 10$~GeV/c. Several observations indicate
that at $p_T>6$ GeV/c, hard processes may dominate particle
production. Below this, measurements of the $B/M$ ratios, the
$\overline{B}/B$ ratios, $R_{CP}$, and $v_2$ indicate that processes
beyond the reach of perturbative QCD are prevalent. Measurements favor
pictures involving either a large contribution from coalescence
processes during hadronization or the presence baryon junctions. The
systematics of the baryon enhancement---including the
$\sqrt{s_{_{NN}}}$ dependence, the collision system dependence
(particularly $d+Au$), and the $B/M$ ratios on the near- and away-side
of trigger hadrons---may be a challenge to the baryon junction
picture. If the coalescence models are correct, identified particle
$v_2$ measurements imply that a highly interacting quark and gluon
phase exists in heavy ion collisions prior to hadron formation. The
existance of such a phase is also favored by the large $v_2$ values
observed for multiply strange hadrons.

\textbf{Acknowledgments:} I'd like to thank the conference organizers
for the invitation to Budapest, along with H. Huang, H-G. Ritter, and
N. Xu for their guidance.

\end{document}